\documentclass[]{aa}

\usepackage{graphicx,times,latexsym,bm}             
\input{epsf.sty}                        
\input{psfig.sty}                       
\begin{document}

   \title{Candidate Milky Way Satellites in the Galactic Halo}
   \author{C. Liu
      \inst{1}\inst{2}\thanks{email: chaoliu@lamost.org}
   \and J. Hu
      \inst{1}
   \and H. Newberg
      \inst{3}
   \and Y. Zhao
      \inst{1}
      }

   \institute{National Astronomical Observatories, Chinese Academy of Sciences, Beijing 100012, China
             \and
             Graduate University of Chinese Academy of Sciences, Beijing 100049, China
             \and
             Department of Physics, Applied Physics, and Astronomy, Rensselaer Polytechnic Institute, Troy NY 12180, USA}

   \date{Received~~; accepted~~}

   \abstract{}{Sloan Digital Sky Survey (SDSS)
    DR5 photometric data with $120^{\circ} < \alpha < 270^{\circ}$, $25^{\circ} < \delta < 70^{\circ}$ are searched
    for new Milky Way companions or substructures in the Galactic halo.}{
    Five candidates are identified as overdensities of faint stellar sources that
     have color-magnitude diagrams similar to those of known globular
     clusters or dwarf spheroidal galaxies.  The distance to each candidate is
     estimated by fitting suitable stellar isochrones to the color-
     magnitude diagrams. Geometric properties and absolute magnitudes are roughly
     measured and used to determine whether the candidates are likely dwarf
     spheroidal galaxies, stellar clusters or tidal debris.}{SDSSJ1000+5730 and SDSSJ1329+2841 are
     likely faint dwarf galaxy candidates while SDSSJ0814+5105,  SDSSJ0821+5608 and SDSSJ1058+2843
      are likely extremely faint
     globular clusters. }
     {Follow-up study is needed to confirm these candidates.
   }

   \authorrunning{C. Liu et al.}
   \titlerunning{Milky Way Satellites Candidates}
\maketitle

\section{Introduction}
\label{sect:intro}

The discovery of tidal streams in the Milky Way from the accretion
of smaller galaxies supports hierarchical merging
cosmogonies(\cite{ll95}). The large-scale tidal arms of the
Sagittarius dwarf spheroidal galaxy accretion event were the first
to be discovered (\cite{yan03}; \cite{Maj03} and \cite{Roc04}), and
more tidal streams continue to be detected in Sloan Digital Sky
Survey (SDSS; \cite{yok00}) data.  The remarkably strong tidal tails
of Palomar 5 are discovered by \cite{ode01}, \cite{Roc02} and
\cite{gd06b}. Another tidal stream is connected with NGC5466
(\cite{bel06a0}; \cite{gj06d}). \cite{bel06a} reveals the so-called
\emph{Field of Streams}, which shows not only Sagittarius arms but
also the Virgo overdensity (\cite{jur05}), Monoceros Ring
(\cite{new02}), Orphan stream(\cite{bel06a,g06}) and a
$60^{\circ}$-long stream between Ursa Major and Sextans
(\cite{gd06c}). The fact that tidal streams are often associated
with globular clusters or dwarf spheroidal galaxies implies that it
is possible to search for tidal streams by identifying new faint
satellite dwarf galaxies, globular clusters or tidal debris and
ascertaining their connections.

New satellite companions of the Milky Way have been discovered in
the recent years. About 150 globular clusters (\cite{har96}) and 9
widely-accepted dwarf spheroidal galaxies (dSph) (\cite{bel06c})
were known before the SDSS. Since 2005, tremendous progress has been
made using SDSS data; at least nine new companions were discovered
within two years. Seven of them are dSph satellites: Ursa Major I
(\cite{wil05a}), Canes Venatici I(\cite{zuc06a}), Bo\"otes
(\cite{bel06b}), Ursa Major II (\cite{zuc06b}; \cite{g06}), Coma
Berenics, Canes Venatici II, Hercules and Leo IV (\cite{bel06c}).
Two of them are probably globular clusters: Willman 1
(\cite{wil05b}) and Segue 1 (\cite{bel06c}).

It is impossible to see these objects in SDSS images since they are
resolved into field stars.  They are discovered by detecting local
stellar overdensities in SDSS stellar databases (\cite{bel06c}) by
adding constraints on magnitude and color index. Overdensity study
is a quick data mining method to find previously unknown satellite
galaxies or globular clusters. We use a method similar to that of
\cite{bel06b}, but utilize different parameters, to scan an area of
the sky bounded by $120^{\circ} < \alpha < 270^{\circ}$, $25^{\circ}
< \delta < 70^{\circ}$, and observe five interesting overdensities
with color-magnitude diagrams similar with a globular cluster or a
dSph. They are candidate globular clusters or dSphs or tidal debris.

The organization of this paper is as follows.  Section 2
describes our method of searching for candidates and how we measure
their properties. Data acquisition and reduction processes, including
the method for identifying candidates, are discussed in section 2.1. A
template matching algorithm for estimating the distance to each overdensity is
described in section 2.2. A rough measurement of geometric
properties and radial profiles are addressed in section 2.3. In
section 3 we discuss the nature of each overdensity. A short
conclusion is included in the last session.

\section{Discovery}
\label{sect:dcv}

\subsection{Data Acquisition and Reduction}
SDSS DR5 provides photometric data in $u$, $g$, $r$, $i$, and $z$
passbands (\cite{Fuk96}), and covers 8000 square degrees of sky
(\cite{ade07}). We select only the objects with stellar profiles
from SDSS databases. SDSS recognizes all point sources as stars,
including quasars and faint galaxies. Thus, galaxy clusters and arms
and halos of bright galaxies contaminate our sample of overdensities.
In order to decrease the frequency of these
contaminants, we use only sources with magnitude $19 < i < 22$ and
color $0 < g-i < 1$. We limited the area of sky searched to avoid
the Sagittarius dwarf spheroidal galaxy's tidal arms. Approximately
$3.6\times10^{7}$ point sources were obtained in the region
$120^{\circ} < \alpha < 270^{\circ}$, $25^{\circ} < \delta <
70^{\circ}$
 from SDSS Casjobs\footnote[1]{http://casjobs.sdss.org/dr5/en/}.
 The magnitudes in each band were corrected for extinction
 using values provided by the SDSS database, which are computed
 following \cite{sch98}.

Data analysis procedures are based on the China Virtual
Observatory's experimental data access service(VO-DAS). We compute
the star counts for this data in bins of size
$0.2^{\circ}\times0.2^{\circ}$ Hence, $750\times225$ bins are
generated.  Field mean density and standard deviation for each bin
are defined by the star counts in the surrounding bins, within a
$11\times11$ window. The standard deviation is derived from all bins
in the box except the center bin.  The strength of the peak in the
center bin is estimated from:

\begin{equation}
\tilde{n}_{center}=(n_{center}-{\mu})/\sigma,\label{eq:9}
\end{equation}

where $n_{center}$ is the number of stars in the center bin, $\mu$
is the average number of stars per bin in the surrounding bins, and
$\sigma$ is the standard deviation.  For most objects, the bin size
is large enough to contain an entire star cluster; the surrounding
bins contain many Galactic stars.  For our sample,
$\tilde{n}_{center}$ is normally distributed with $\tilde{\mu}=0.33$
and $\tilde{\sigma}=0.56$.  We selected all peaks with
$\tilde{n}_{center}>\boldsymbol{\tilde{\mu}+3\tilde{\sigma}\sim2}$.
From these statistics, we expect about $0.27\%$ of the bins we
search will contain statistical fluctuations high enough to be
detected as a peak. We searched 168750 bins, of which we excluded
18627 zero bins (sky areas which are not covered by SDSS).
Therefore, we expect about 405 overdensities are actually caused by
field star density fluctuation.

We actually identified 524 overdensity bins, which is 119 more than
would be expected from statistics.  The 524 overdensities include
statistical fluctuations, previously identified Milky Way
satellites, bright galaxies, bright stars, and galaxy clusters, in
addition to the new Milky Way satellites we would like to find.  We
use additional information from the color-magnitude distribution of
stars in the overdensities to cull out the statistical fluctuations,
as described below.  Known objects including bright galaxies, galaxy
clusters, and bright stars are manually identified and eliminated.
The known Milky Way satellites that were re-discovered by this
technique include: UMa II, UMa I, Willman 1, Pal4, CVn II, CVn I,
NGC5272, NGC5466, NGC6205, NGC6341, Draco, and Leo III, some of
which cover more than 2 bins due to their large angular sizes.
Overdensity bins which are close to the survey boundaries (within
$0.5^{\circ}$) are also ignored.

Density contour diagrams and CMDs were generated within a
$1^{\circ}\times1^{\circ}$ area centered at the remaining
overdensity bins. Candidates were manually identified by comparing
the density contour diagram and CMD of each overdensities with those
of 9 known objects: Pal 4, UMa I, UMa II, CVn I, Draco, NGC 6341,
NGC 5466, Willman 1 and NGC 6205. The distance modulus (DM) of these
templates ranges from 14.2mag to 21.7mag. This ensures that the
templates cover all possible morphological features for a CMD,
{since the stellar populations that are probed within our magnitude
limits depend on the distance modulus.}

\begin{figure*}
   \centering
   \resizebox{0.7\linewidth}{!}
   {\includegraphics[]{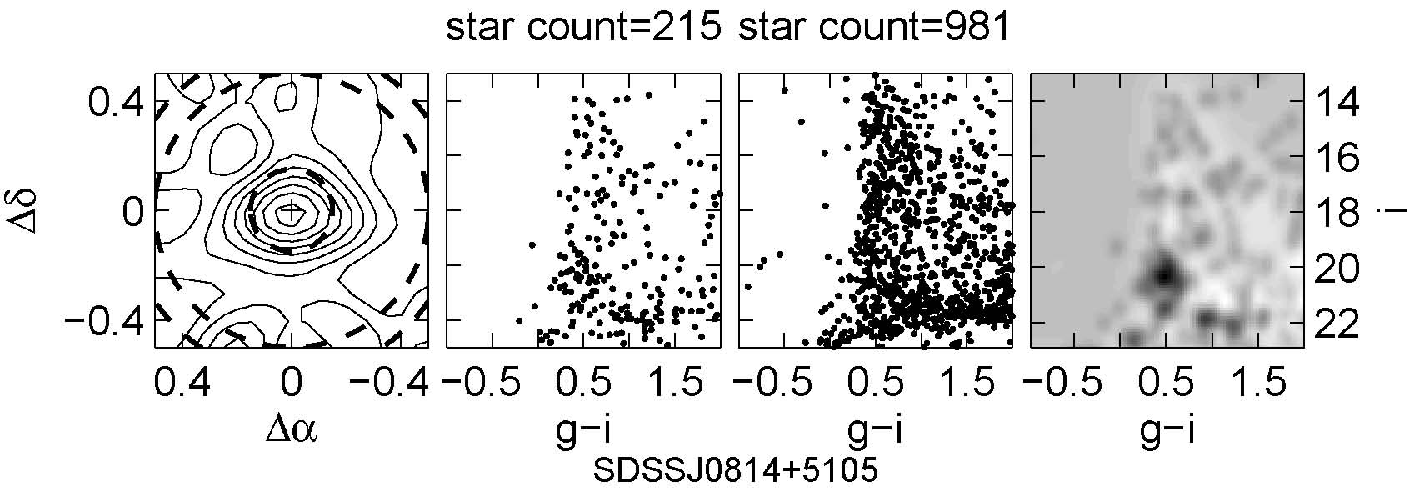}}\\
\resizebox{0.7\linewidth}{!}
   {\includegraphics[]{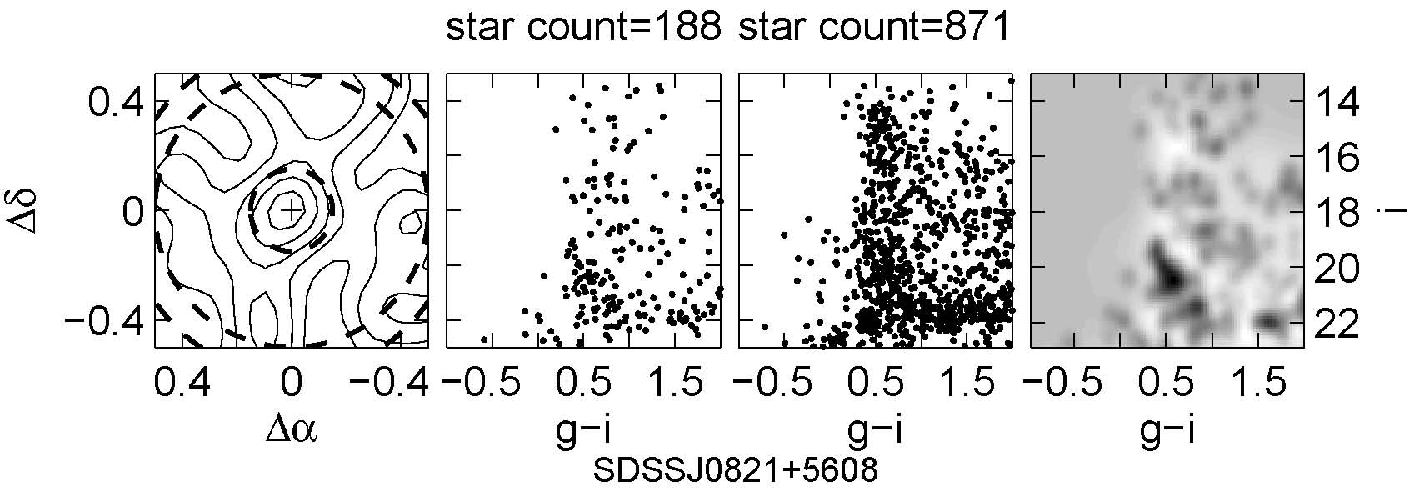}}\\
   \resizebox{0.7\linewidth}{!}
   {\includegraphics[]{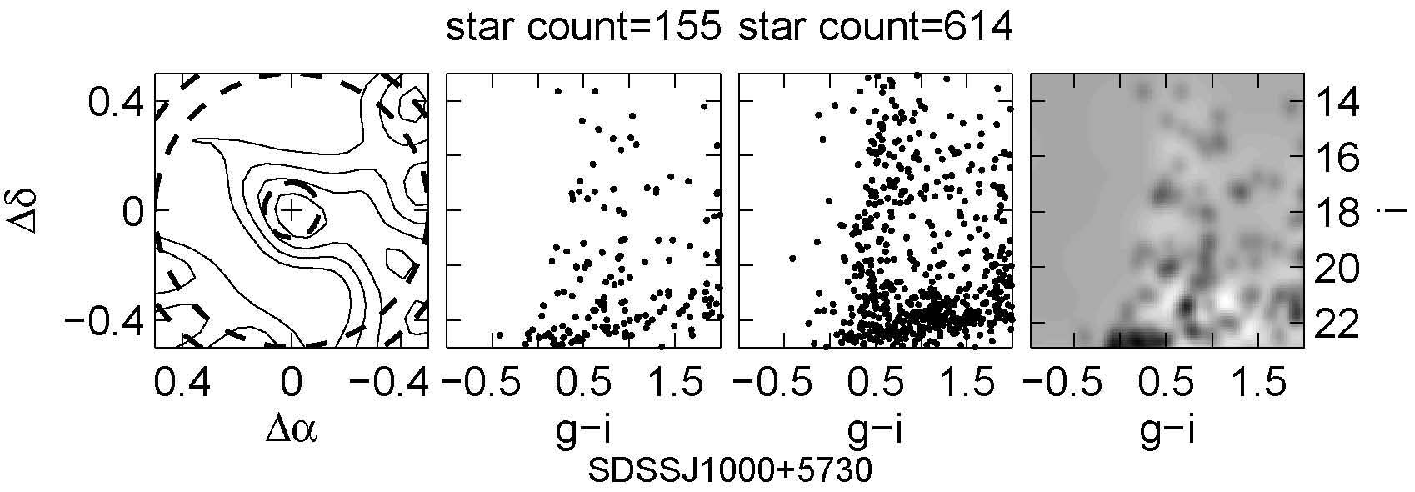}}\\
   \resizebox{0.7\linewidth}{!}
   {\includegraphics[]{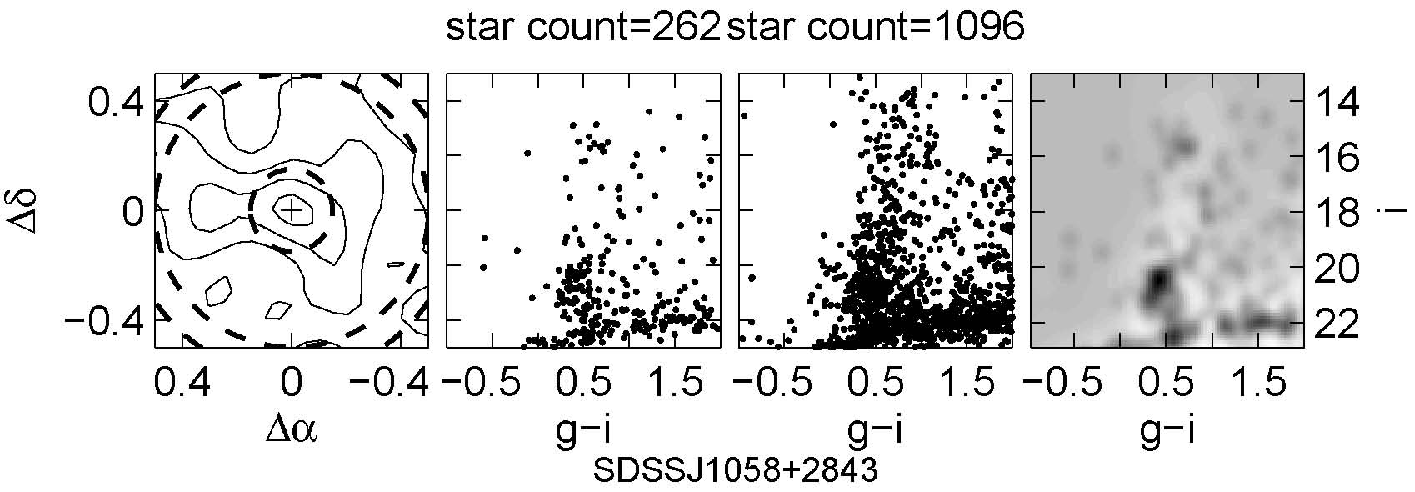}}\\
   \resizebox{0.7\linewidth}{!}
   {\includegraphics[]{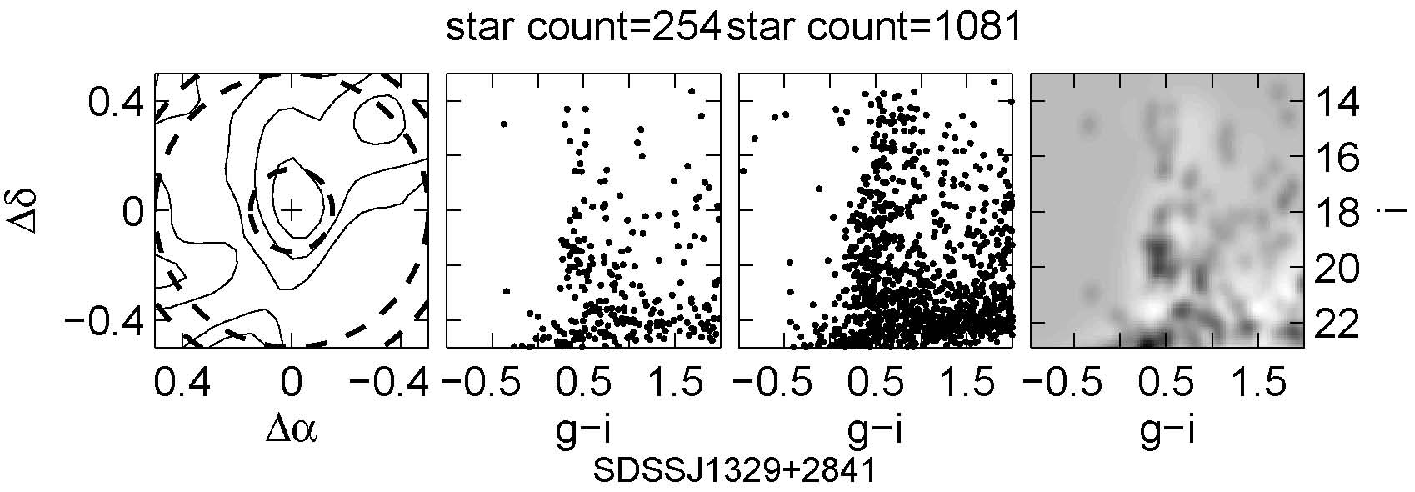}}\\
   \caption{\label{Fig:fig1}
   Overdensity detections.
   The first column of diagrams are density contours.  Each density contour is
   constructed by nonparametric
   distribution density estimation with a Gaussian kernel and a window width of
   $0.1^{\circ}$, using stars with $19<i<22$mag
   located in the $1^{\circ}\times1^{\circ}$ area around the candidate.
   The second column of diagrams displays the core CMDs of overdensities.
   Points in this column are samples from
   the smallest circle ($r=0.1^{\circ}$ or $0.15^{\circ}$)
   overlaid on density contours. The third column of diagrams displays
   field CMDs of the overdensities. Points in this column are samples
   from the annuli between the intermediate circle ($r=0.5^{\circ}$) and the
   biggest circle ($r=0.6^{\circ}$). The right most column of diagrams
   are Hess diagrams which are the result of subtraction between the
   core CMDs and corresponding field CMDs with normalization on
    total star counts. The size of each bin in the
   Hess diagrams is $0.03 mag \times 0.1 mag$.}
   \end{figure*}

Five overdensities are finally chosen from over 500 initial
candidates. Figure \ref{Fig:fig1} displays density contours, core
(circle with $r=0.1^{\circ}$ or $r=0.15^{\circ}$) CMDs, field
(annuli area between $r=0.5^{\circ}$ and $r=0.6^{\circ}$) CMDs and
Hess diagrams (subtraction between corresponding center area CMDs
and field CMDs with normalized star counts) for these five candidate
Milky Way satellites.

\subsection{Distances to the Overdensities}

Distances to the five candidates are estimated by fitting isochrone
lines to the CMD of each overdensity. An automatic template-matching
algorithm was developed to discover the best-fit Padova isochrone
lines(\cite{gir04}) for each candidate. Stellar evolution isochrone
lines are converted to masks where the color index $g-i$ changes
from -1 to 2 with step of 0.03 and the magnitude $i$ changes from
-10 to 10 with step of 0.1. Thus a binary $100\times200$ image mask
for each isochrone line is formed. The value of each pixel in the
image is either 1 (in the neighborhood of an isochrone line) or 0
(everywhere else).  A series of mask templates are created for a
range of metallicity, age and distance moduli. For each candidate,
the template which minimizes $R$, as defined by the following
formula, is selected as the best match.
\begin{equation}
R=\frac{\sum\limits_{g-i,i}{(C(g-i,i)(1-T(g-i,i,Z,A,DM)))}}{\sum\limits_{g-i,i}{C(g-i,i)}}
\label{eq:1}
\end{equation}
where $C(g-i,i)$ is the candidate's Hess diagram ($i=13-22$mag).
$T(g-i,i,Z,A,DM)$ is a template mask with specific metallicity $Z$,
age $A$ and distance modulus $DM$. The value of $Z$ is 0.0001,
0.0004, 0.001, or 0.004. The value of $A$, in $log_{10}(yr)$, ranges
from 9.5 to 10.25 with a step of 0.05. And the value of $DM$ ranges
from 13mag to 21.9mag with a step of 0.1mag.  In total 5760
templates were evaluated for each overdensity. The value of R also
measures the goodness of the fit.  The smaller R is, the better the
fit.

\begin{figure*}
    \centering \resizebox{0.5\linewidth}{!}
    {\includegraphics[]{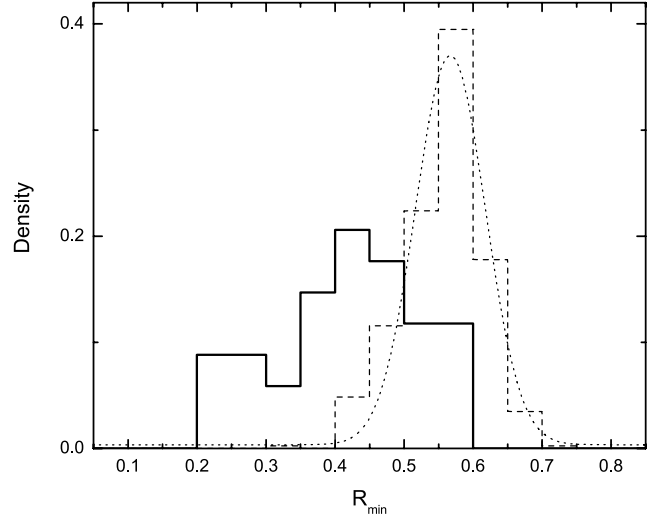}}
\caption{\label{Fig:Rmin} Comparison of $R_{min}$ for real objects
compared with all original two sigma detections.  Two populations
of $R_{min}$ values distribution are displayed. The solid stair line
shows all known
objects, including globular clusters, dwarf galaxies, bright galaxies
and galaxy clusters, and our five candidates. The distribution shows
two peaks.  The peak between 0.2 and 0.3 is due to 5
known globular clusters.  The peak at 0.4 is associated
with dwarf galaxies and normal galaxies. As comparison, the dashed
stair line shows the distribution of all other overdensities
with a gaussian fitting curve(dot line), where the average value is
$R_{min}=0.57$ and $\sigma=0.1$.  This figure shows that the globular
clusters have the smallest $R_{min}$, the dwarf galaxies
reach a larger radus, and both objects have a significantly smaller
$R_{min}$ than non-candidates.}
\end{figure*}

$R_{min}$, the minimum value of $R$, measures how many stars, in
percent, are unrelated to the best fit isochrone line in the Hess
diagrams. In the extreme case that stars are uniformly distributed
in color-magnitude plane, considering that the isochrone line is
only a thin line and cannot cover many stars in the plane,
$R_{min}\sim1$ is derived. In the opposite extreme case that the
candidate contains no background stars so that its features are
completely described by a template, $R_{min}\sim0$ is reasonable.
$R_{min}$ distributions for all known objects, candidates
satellites and other overdensities are shown in Figure
\ref{Fig:Rmin}. The population of known objects and candidates
addresses significantly smaller $R_{min}$ values than that of
non-candidates overdensities.

Specifically, we use 9 known globular clusters and dwarf spheroidal
galaxies: UMa I \& II, CVn I \& II, Pal4, NGC6205, NGC5466, NGC5272,
and NGC6341, to test the effectiveness of the algorithm. They are
all located in the detected area and were all detected by our
selection process. Their distance moduli range from 14 to 22, as
measured by previous authors (which are listed in Table
\ref{Tab:tab2}).
The standard deviation in the accuracy of the DM is $\sigma=0.23$mag.

\begin{figure*}
  \centering \resizebox{0.8\linewidth}{!}
  {\includegraphics[]{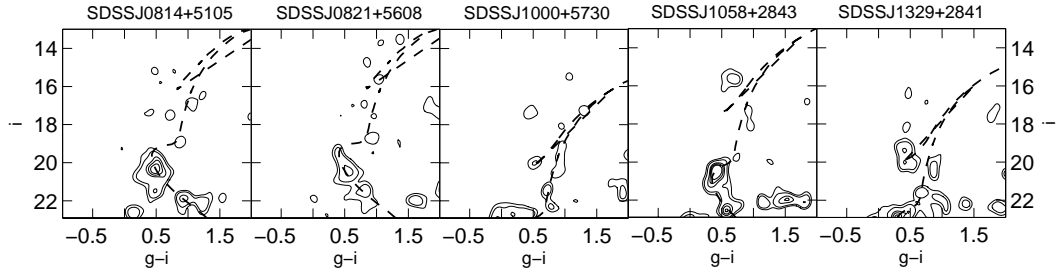}}
\caption{\label{Fig:fig3} The Hess contours of candidates with the
best-fitting isochrone lines overlaid on them. Contours
 are positive densities (0.05, 0.1, 0.15, 0.2, 0.25,
0.3 stars per bin). The size of each bin is $0.03 mag \times 0.1
mag$. Best-fitting stellar evolution isochrone lines are overlaid
with dashed lines.}
\end{figure*}

Best fit isochrone lines are shown together with Hess diagrams of
our candidate Milky Way satellites in Figure \ref{Fig:fig3}.  The
distance modulus of the best fit template is an estimate of the
distance to each overdensity.  Table \ref{Tab:tab0} displays the
parameters of the matching templates for each candidates, though we
don't expect that this method of isochrone fitting will produce
accurate measurements of the ages and metallicities of each
candidate. We list our best fit metallicity and age parameters
simply to show that they are not unreasonable for star clusters or
dwarf galaxies in the Galactic halo.

$R_{min}$ values in table \ref{Tab:tab2} again show the same trend
with Figure \ref{Fig:Rmin} that globular clusters have smaller
$R_{min}$ values, while low surface brightness dSphs (for example
UMa II) have larger $R_{min}$ values. The fact that an acceptable DM
is derived for UMa II, although it has a higher $R_{min}$ that
reaches the non-candidates range(see Figure \ref{Fig:Rmin}), gives
us confidence that our technique for finding distances is valid even
for these low surface brightness candidates.

\subsection{Properties of Candidates}


Center positions, geometric properties and absolute magnitudes are
estimated for each candidate and listed in Table \ref{Tab:tab0}.

\begin{figure*}
  \centering \resizebox{0.8\linewidth}{!}
  {\includegraphics[]{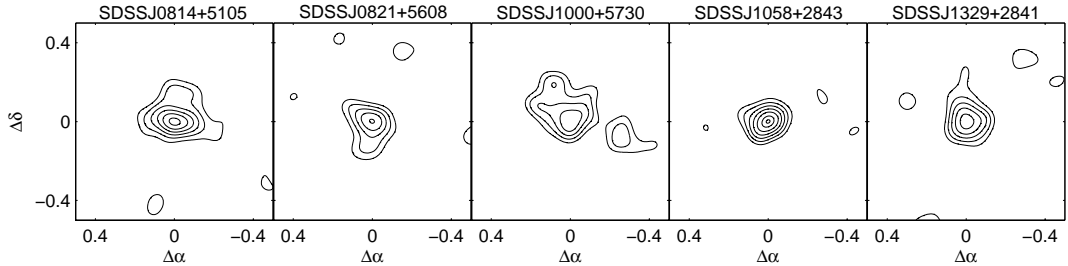}}
\caption{\label{Fig:fig32} Isodensity contours of five candidates.
The contour levels are 2, 3, 4, 5, 6, 7 and 7.5$\sigma$ above the
background. The samples that generate the contours are selected from the
corresponding Hess diagram.}
\end{figure*}

In order to measure position and geometric properties for each
candidate Milky Way satellite, stars associated with candidates are
selected.  We select stars which are located within the
$1^{\circ}\times1^{\circ}$ around the center position of candidates
and also within bins in the Hess diagram (right panel in Figure
\ref{Fig:fig1}) that have more than 0.05 stars per bin.
%
Density contour diagrams of the five candidates are derived by
counting selected stars and are displayed in Figure \ref{Fig:fig32}.

%

Because there are relatively few stars in our samples, we did not
fit for the ellipticity of the candidates.  We fit only circularly
symmetric profiles. Background level is estimated by averaging star
density of an annulus area around each candidates with radius from
30arcmin to 60arcmin. It is then subtracted from radial profile
before fitting.

\begin{figure*}
\begin{minipage}[t]{0.2\linewidth}
  \centering \resizebox{1\linewidth}{!}
  {\includegraphics[]{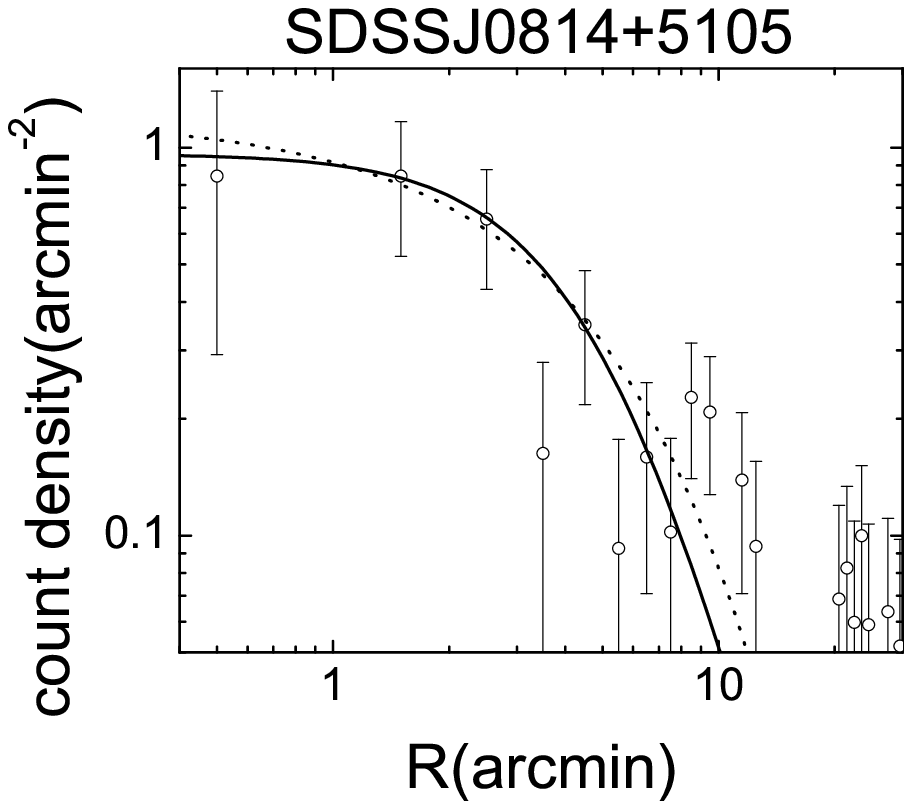}}
  \end{minipage}%
  \begin{minipage}[t]{0.2\linewidth}
  \centering \resizebox{1\linewidth}{!}
  {\includegraphics[]{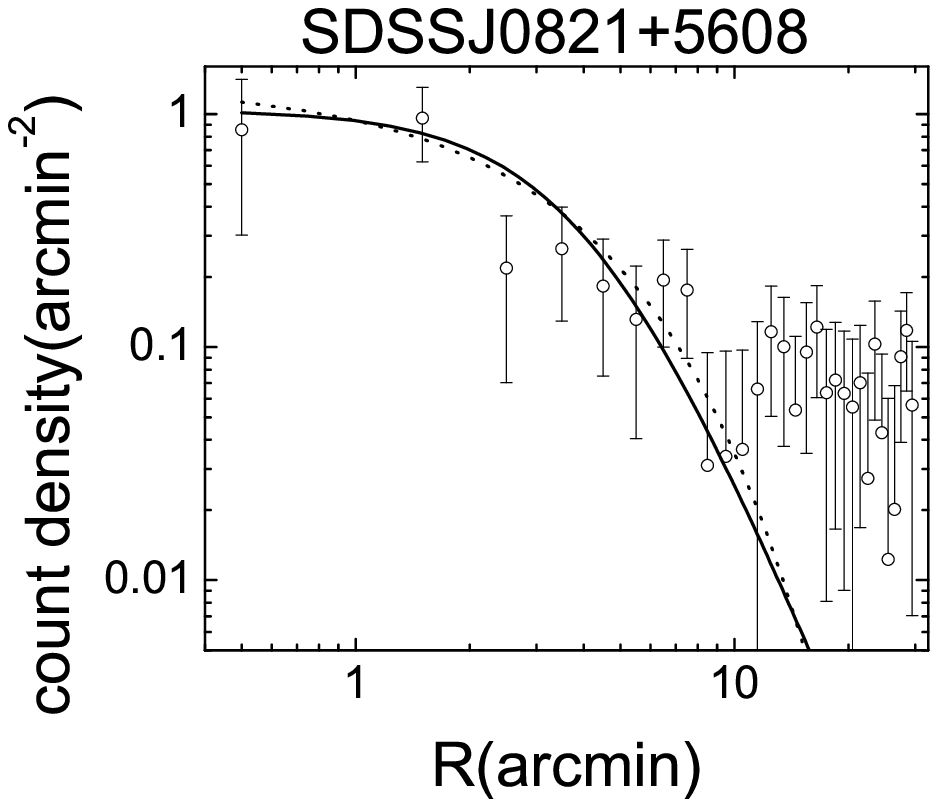}}
  SDSSJ0821+5608
  \end{minipage}%
  \begin{minipage}[t]{0.2\linewidth}
  \centering \resizebox{1\linewidth}{!}
  {\includegraphics[]{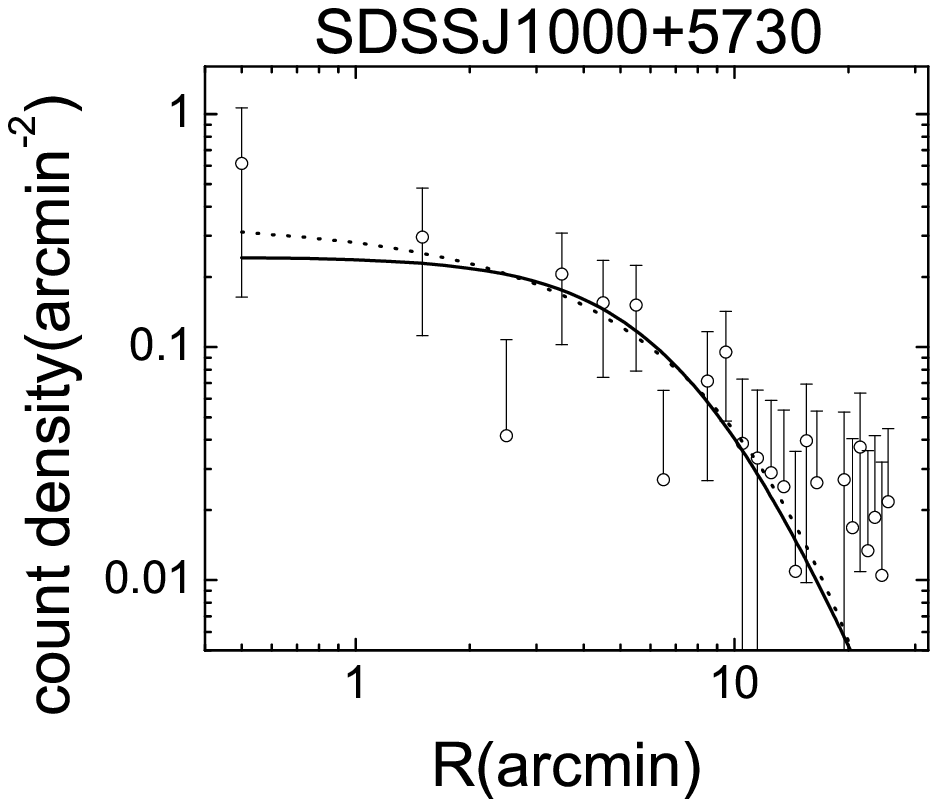}}
  SDSSJ1000+5730
  \end{minipage}%
  \begin{minipage}[t]{0.2\linewidth}
  \centering \resizebox{1\linewidth}{!}
  {\includegraphics[]{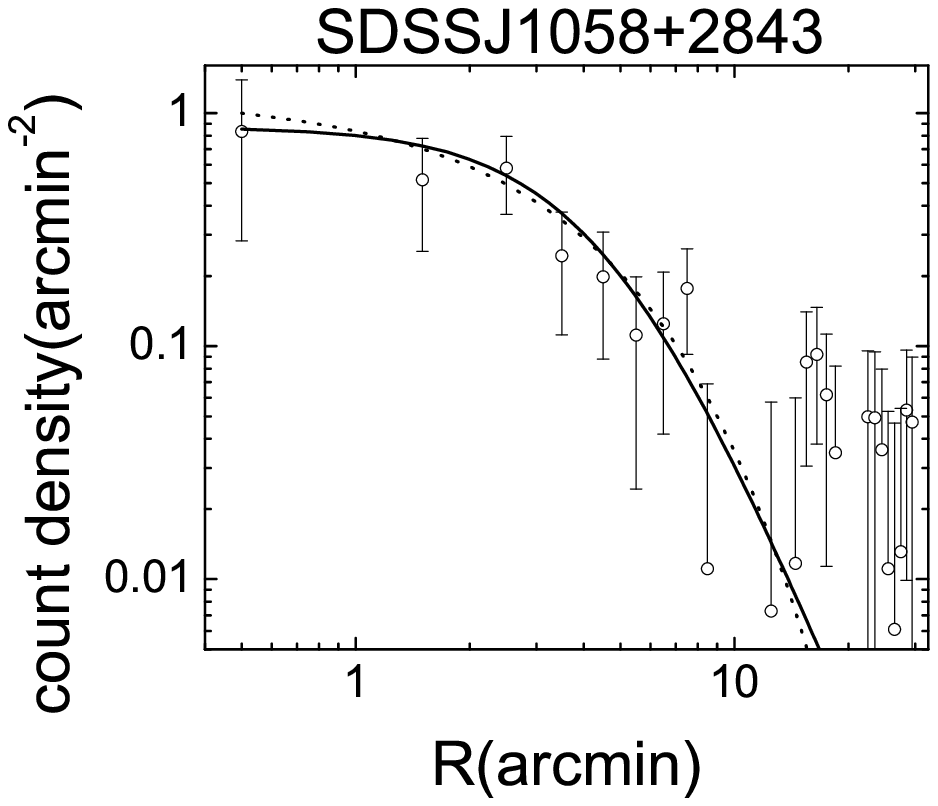}}
  SDSSJ1058+2843
  \end{minipage}%
  \begin{minipage}[t]{0.2\linewidth}
  \centering \resizebox{1\linewidth}{!}
  {\includegraphics[]{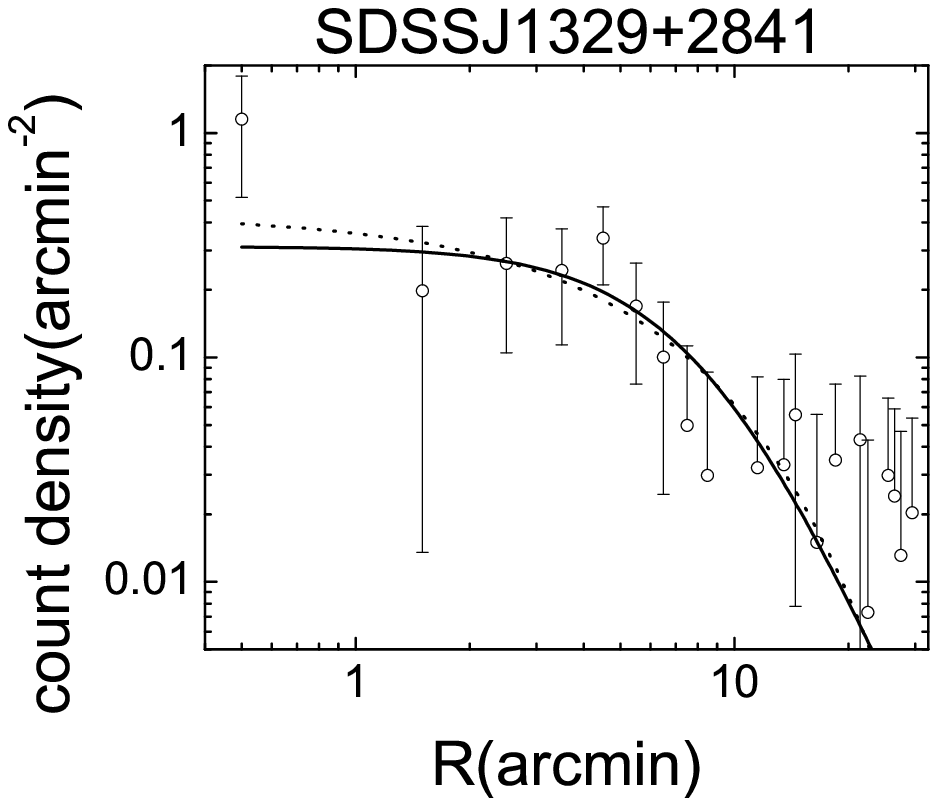}}
  SDSSJ1329+2841
\end{minipage}%
  \caption{\label{Fig:fig4}Density radial profiles for
  candidates.\textbf{Background levels are subtracted before fitting models.}
  The solid lines are Plummer model fit curves. The dashed lines are exponential model fit curves.}
\end{figure*}

In order to check our methods, we compare the half-light radius of
Pal5 and Boo using our method and compare the results with previous
measurements in the literature. For Pal5 we derive
$r_{h}=2.64^{\prime}\pm{0.07}$ for the exponential model and
$r_{h}=2.91^{\prime}\pm{0.07}$ for the Plummer model. These values
are similar to the previously measured value of
$r_{h}=2.96^{\prime}$ in \cite{har96}. For Boo, we find
$r_{h}=14.4^{\prime}\pm{1.8}$ for the exponential model and
$r_{h}=14.6^{\prime}\pm{1.5}$ for the Plummer model, while in
\cite{bel06b} corresponding values are $r_{h}=13.0^{\prime}\pm{0.7}$
and $r_{h}=12.6^{\prime}\pm{0.7}$. The results support our methods
for estimating globular cluster and dwarf galaxy parameters, though
there may be larger errors for lower luminosity objects.
All candidates are fitted by exponential and Plummer
models(\cite{MI06}). Radial profiles and fitting curves are
displayed in Figure \ref{Fig:fig4}. Half-light radii derived by
integrating exponential and Plummer profiles are tabulated in Table
\ref{Tab:tab0}. Geometrical sizes of these candidates are computed
from the estimated distances and the angular sizes from each model
profile and also show in Table \ref{Tab:tab0}. 

To estimate absolute magnitude of candidates we use Hess diagrams
derived by subtracting field Hess diagrams from those generated by
stars inside $\rm r_{h}$, normalized by area. Overlapping
corresponding best matched isochrone masks on these Hess diagrams,
only star counts located at neighbors of isochrone lines are kept.
Then, g-band and r-band magnitudes are computed by integrating all
positive flux values in these masked Hess diagrams. Total color
index $\rm (B-V)_{tot}$ and Total apparent magnitude $\rm V{tot}$
are conducted from Formula \ref{eq:6} and \ref{eq:7} (\cite{Fuk96}).
\begin{equation}
(B-V)_{tot}=(g_{tot}-r_{tot}+0.23)/1.05\label{eq:6}
\end{equation}
\begin{equation}
V_{tot}=r_{tot}+0.49(B-V)_{tot}-0.11\label{eq:7}
\end{equation}
\begin{equation}
M_{V,tot}=V_{tot}+5-5log(d)\label{eq:8}
\end{equation}
Consequently, absolute magnitudes are computed via \ref{eq:8}.
Absolute magnitudes estimated with different radial profile models
are tabulated in Table \ref{Tab:tab0}. As a test, we estimate the
absolute magnitudes of Com, CVn II, Her and Leo IV, which are low
luminosity dwarf galaxies discovered in SDSS data(\cite{bel06c}),
using this method. The maximum bias of all four objects between our
results and the literature's is $\rm |\Delta M_{V,tot}|=0.55mag$,
which is smaller than the standard error(0.6mag) mentioned in the
literature. ANOVA analysis also shows that the estimation of our
method is not significantly different with those from \cite{bel06c}.

\section{Discussion}
\label{sect:disc}

It is difficult to identify all 5 candidates reliably using only
SDSS data; accurate follow up observations using a large telescope
is required to determine the types, which could be: star clusters,
dwarf spheroidal galaxies, tidal debris or chance superposition of
field stars.  We expect that these candidates are not merely field
stars.

\begin{figure*}
\centering
   \resizebox{0.7\linewidth}{!}
   {\includegraphics[]{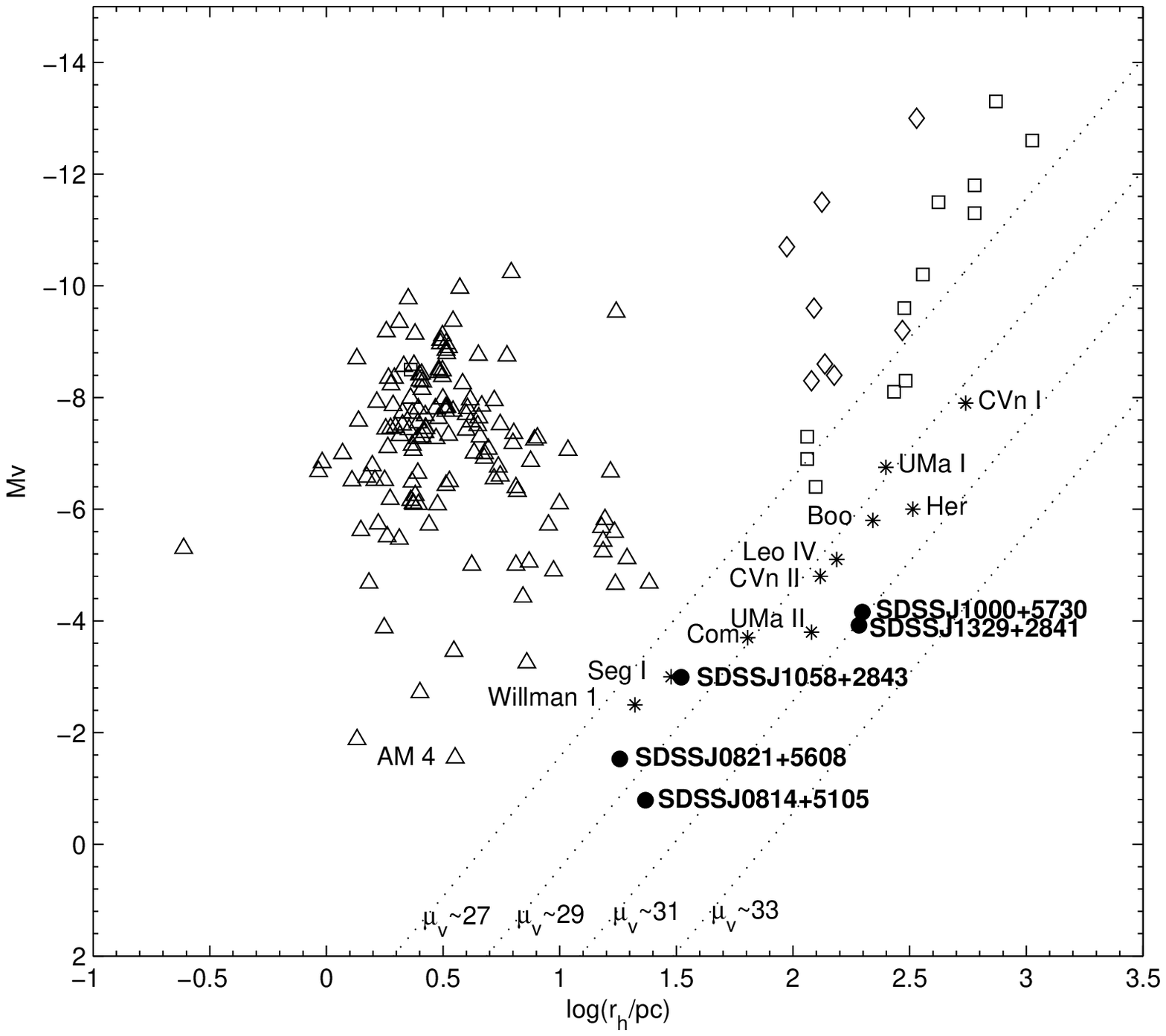}}
   \caption{\label{Fig:fig7}Location of different
   classes of objects in the plane of half light
   radius vs. absolute magnitude. Lines of constant surface
   brightness are marked. Filled circles with bold labels are the 5 candidates
   revealed in this paper. Triangles are known Globular Clusters
   (\cite{har96}). Stars with labels are Milky Way
satellites discovered from SDSS(\cite{bel06b}, \cite{bel06c},
\cite{zuc06b}, \cite{wil05a} and \cite{wil05b}), Diamonds are dwarf
galaxies in the local group(\cite{ih95} and \cite{Mat98}).
Rectangles are Andromeda dwarf galaxies(\cite{MI06} and
\cite{Mat06})}

\end{figure*}

\cite{bel06c} uses the $R_{h}$ vs. $M_{V}$ plane to identify Her,
Leo IV, CVn II, Com as dSphs, while Seg I is an unusual low
luminosity globular cluster. We adopt that same method, and plot our
candidates with known Milky Way satellites in Figure \ref{Fig:fig7}.
The half-light radii of 5 candidates are derived by integrating
exponential profiles and Plummer profiles. The absolute magnitudes
are derived by adding all possible member stars flux which located
in half-light radius and time 2. We plot the 5 candidates in $R_{h}$
vs. $M_{V}$ plane with the mean values derived by exponential and
Plummer model. $R_{h}$ and $M_{V}$ of known globular clusters are
come from \cite{har96}, and those of SDSS discovered Milky Way dwarf
spheroidal galaxies are come from \cite{bel06b}, \cite{bel06c},
\cite{zuc06b}, \cite{wil05a} and \cite{wil05b}, those of dwarf
galaxies in the local group are come from \cite{ih95} and
\cite{Mat98}, and those of Andromeda dwarf galaxies are come from
\cite{MI06} and \cite{Mat06}.

\begin{figure*}
\centering
   \begin{minipage}[t]{0.33\linewidth}
  \centering \resizebox{1\linewidth}{!}
  {\includegraphics[]{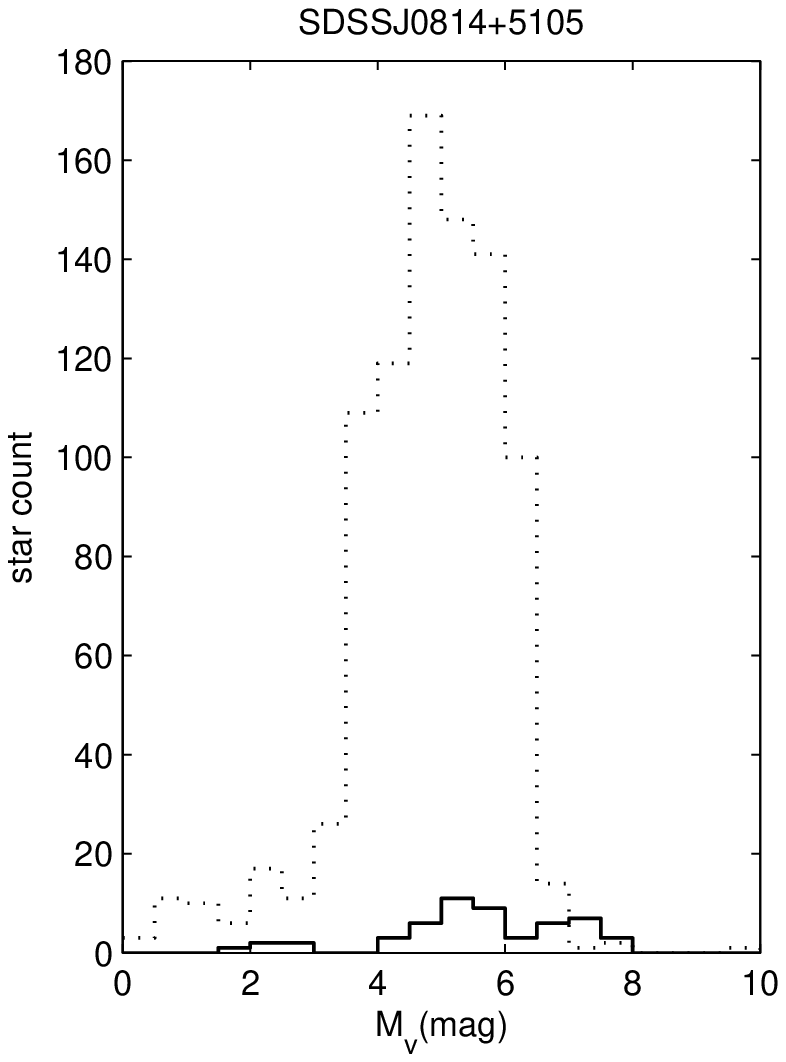}}
  \end{minipage}%
  \begin{minipage}[t]{0.33\linewidth}
   \centering \resizebox{1\linewidth}{!}
  {\includegraphics[]{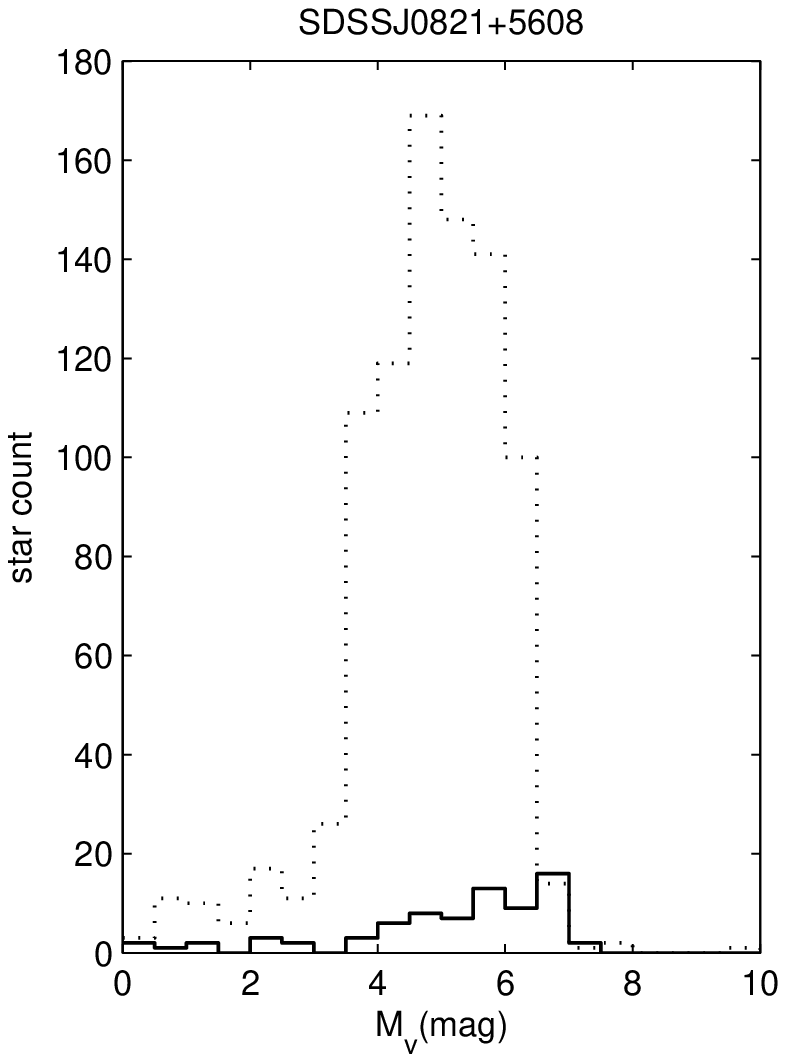}}
  \end{minipage}%
  \begin{minipage}[t]{0.33\linewidth}
  \centering \resizebox{1\linewidth}{!}
  {\includegraphics[]{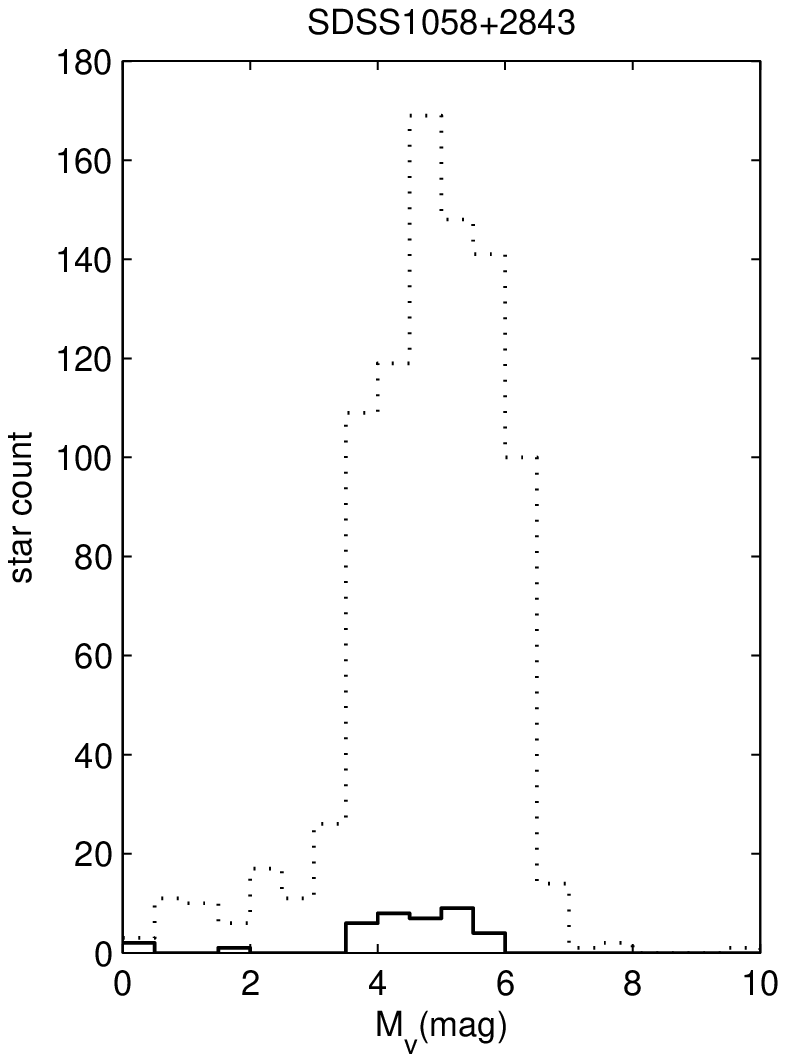}}
  \end{minipage}%
   \caption{\label{Fig:fig8} Luminosity function comparison between
   SDSSJ0814+5105, SDSSJ0821+5608 and SDSSJ1058+2843, marked as solid line, and Pal 5, marked as dot
   line, within half-light radii. Stars between 4 and 6 mag correspond
   to spectral types from F to G in main sequence.  Star counts of F and
   G stars are used to roughly estimate the total mass.}
   \end{figure*}

\emph{SDSSJ0814+5105, SDSSJ0821+5608 and SDSSJ1058+2843}:

SDSSJ0814+5105, SDSSJ0821+5608 and SDSSJ1058+2843 have surface
brightness which are similar to other known satellites discovered
from SDSS with extremely faint absolute magnitudes, even fainter
than AM4, the faintest one in \cite{har96}. Their location in Figure
\ref{Fig:fig7} suggests that these stellar systems are more similar
to globular clusters than to dwarf galaxies. Tidal radii of
SDSSJ0814+5105 , SDSSJ0821+5608 and SDSSJ1058+2843 in Figure
\ref{Fig:fig7} are computed using the equation
%
\begin{equation}
r_{tidal}\sim{R(\frac{M_{cand}}{3M_{MW}})^{1/3}} \label{eq:3}
\end{equation}
 from \cite{bt87}, where $R$ is the candidate's galactocentric
 distance, $M_{cand}$ is its total mass, and $M_{MW}$ is the total
 mass of the Milky Way within $R$. We assume that the distance from
 the Sun to the center of the Milky Way is 8kpc and
 $v_{c}=220kms^{-1}$ at distance $R$. $M_{cand}$ is estimated by
 comparing star counts of F and G stars located in main sequences of
 the three candidates and Pal 5 (Figure \ref{Fig:fig8}). Since
the formula for the tidal radius does not include dark matter, the
actual tidal radius is larger than estimated if the candidates are
 dwarf galaxies with substantial dark matter content.

%

\begin{figure*}

   \centering
   \resizebox{0.7\linewidth}{!}
   {\includegraphics[]{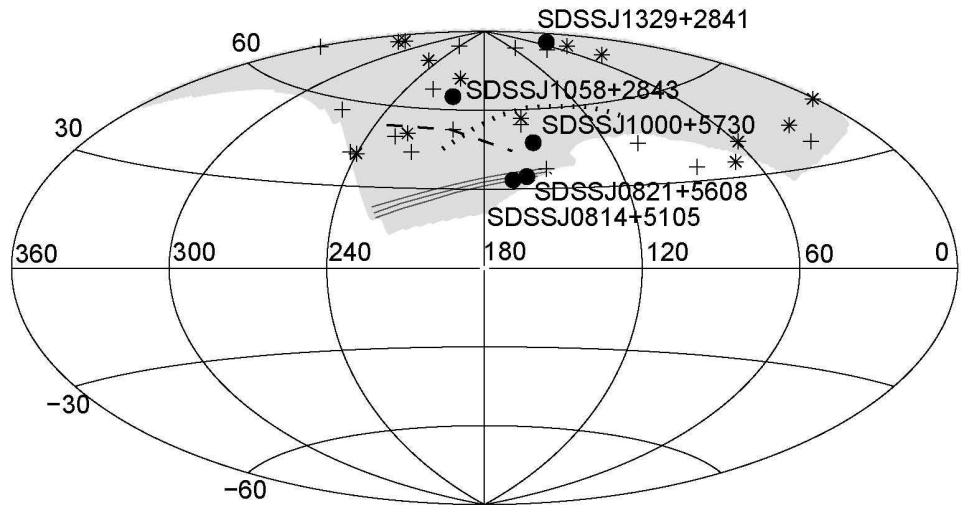}}

   \caption{\label{Fig:fig6} Overdensities' positions in galactic coordinates in Aitoff projection.
    The gray area encloses the main area of North Galactic Cap covered by the SDSS DR5 data.
    Filled circles are candidates in this paper. Plus symbols are UMa II, Willman 1,
   UMa I, CVn I, CVn II, Her, Leo IV, Com and Seg 1. Star symbols are globular clusters
   in the SDSS DR5 area we detect. The triple thin solid lines are Anticenter Stream of \cite{g06c} which
   may be related to the Monoceros Ring. The thin dotted line is GD-1.
   The thin dashed line is Orphan stream. }
\end{figure*}

Notice that SDSSJ0814+5105 and SDSSJ0821+5608 overlap the Anticenter
Stream (\cite{g06c}), which may be related to Monoceros Ring
displayed in Figure \ref{Fig:fig6}. According to \cite{new02} the
turnoff magnitude of Monoceros Ring is at $g=19.4$. In Figure
\ref{Fig:fig1} we estimate SDSSJ0814+5105 and SDSSJ0821+5608 has a
magnitude of the turnoff as $g=19.5$. They are almost at the same
distance. However, according to \cite{g06c}, the Anticenter Stream
has a turnoff magnitude as $g=18.7$, quite different from the
results of \cite{new02} and substantially closer than the new
candidates. The question is, are they globular clusters in the
Monoceros stream, or are they just part of the debris more compact
than in other areas?

\emph{SDSSJ1000+5730 \& SDSSJ1329+2841}: SDSSJ1000+5730 and
SDSSJ1329+2841 are faint, extended objects with half light radii
similar to that of most of dSphs.  They have surface brightness near
$\mu_{V}\sim{31} mag/arcsec^{2}$, which are lower than all known
dSphs discovered from SDSS. Their type cannot be determined only by
SDSS due to their faintness, but they look like dwarf galaxies.

%

\section{Conclusions}
\label{sect:conclusion}

In this paper we report five interesting overdensities detected in the
SDSS database. They show features of globular clusters or dSphs in
their CMDs. Mass estimation for SDSSJ0814+5105, SDSSJ0821+5608 and
SDSSJ1058+2843 indicate that their tidal radii are bigger than
their half-light radius, which implies that they are likely globular
clusters with low surface brightness. The half-light radius and
absolute magnitude of SDSSJ1329+2841 suggest a likely dSph. Although
surface magnitude measurements suggest that SDSSJ1000+5730 is
fainter than all known dSphs, its Hess diagram implies that it is an
interesting object that needs follow-up observations.

H.N. acknowledges funding from the National Science foundation (AST
07-33161).

Funding for the SDSS and SDSS-II has been provided by the Alfred P.
Sloan Foundation, the Participating Institutions, the National
Science Foundation, the U.S. Department of Energy, the National
Aeronautics and Space Administration, the Japanese Monbukagakusho,
the Max Planck Society, and the Higher Education Funding Council for
England. The SDSS Web Site is http://www.sdss.org/.

The SDSS is managed by the Astrophysical Research Consortium for the
Participating Institutions. The Participating Institutions are the
American Museum of Natural History, Astrophysical Institute Potsdam,
University of Basel, University of Cambridge, Case Western Reserve
University, University of Chicago, Drexel University, Fermilab, the
Institute for Advanced Study, the Japan Participation Group, Johns
Hopkins University, the Joint Institute for Nuclear Astrophysics,
the Kavli Institute for Particle Astrophysics and Cosmology, the
Korean Scientist Group, the Chinese Academy of Sciences (LAMOST),
Los Alamos National Laboratory, the Max-Planck-Institute for
Astronomy (MPIA), the Max-Planck-Institute for Astrophysics (MPA),
New Mexico State University, Ohio State University, University of
Pittsburgh, University of Portsmouth, Princeton University, the
United States Naval Observatory, and the University of Washington.

\clearpage

\begin{table*}[]
\caption[]{Comparison of distances from template matching method with
previous studies} \label{Tab:tab2}
\begin{center}\begin{tabular}{l|cccccl}
  \hline\hline

Name & Z from template & Age from template & DM from template & DM from references & $R_{min}$ & References\\
    & & ($log_{10}(yr)$) &(mag)&(mag)& &\\

 \hline
UMa II & 0.0004 & 10.1 & 17.9 & 17.5 & 0.50 & \cite{zuc06b}\\
UMa I & 0.0001 & 10 & 19.9 & 20 & 0.41 & \cite{wil05a}\\
Pal4 & 0.001 & 10.2 & 20.2 & 20.02 & 0.37 & \cite{har96}\\
CVn II & 0.0004 & 10.2 & 20.9 & 20.9 & 0.25 & \cite{bel06c}\\
CVn I & 0.0004 & 10.15 & 21.7 & 21.75 & 0.28 & \cite{zuc06a}\\
NGC5272 & 0.001 & 10.1 & 15 & 15.04 & 0.21 & \cite{har96}\\
NGC5466 & 0.001 & 10.05 & 16.1 & 16.1 & 0.34 & \cite{har96}\\
NGC6205 & 0.004 & 9.95 & 14.7 & 14.28 & 0.21 & \cite{har96}\\
NGC6341 & 0.001 & 10 & 14.9 & 14.59 & 0.28 & \cite{har96}\\
\hline
\end{tabular}\end{center}
\end{table*}

\begin{table*}[]
\caption[]{Parameters of each overdensity.} \label{Tab:tab0}
\begin{center}\begin{tabular}{c|ccccc}
  \hline\hline

 Parameter & SDSSJ0814+5105 & SDSSJ0821+5608 &
SDSSJ1000+5730 &
SDSSJ1058+2843 & SDSSJ1329+2841\\

 \hline
RA(J2000) & $\rm 08^{h}13^{m}42^{s}$&$\rm 08^{h}21^{m}15^{s}$ &$\rm 10^{h}00^{m}28^{s}$ &$\rm 10^{h}58^{m}04^{s}$ &$\rm 13^{h}29^{m}13^{s}$ \\
DEC(J2000) & $\rm +51^{\circ}05^{\prime}27^{\prime\prime}$&$\rm +56^{\circ}08^{\prime}16^{\prime\prime}$ &$\rm +57^{\circ}30^{\prime}10^{\prime\prime}$ &$\rm +28^{\circ}42^{\prime}39^{\prime\prime}$ &$\rm +28^{\circ}41^{\prime}27^{\prime\prime}$ \\
l(deg) & 167.743& 161.665& 155.506& 202.649& 45.716\\
b(deg) & 33.449& 34.615& 47.372& 64.966& 81.513\\
Z  & 0.004& 0.004& 0.001& 0.004& 0.0004\\
Age($\rm log_{10}(yr)$) & 10.5& 10& 10.15& 9.95& 10.1\\
$\rm (m-M)_{0}$(mag) & 15.7& 15.7& 19.6& 16.9& 19.4\\
$\rm R_{min}$ & 0.5& 0.45& 0.47& 0.36& 0.58\\
Distance(kpc)  & $13.8^{+1.5}_{-1.4}$& $13.8^{+1.5}_{-1.4}$& $83.2^{+9.3}_{-8.4}$& $24^{+2.7}_{-2.4}$& $75.9^{+8.5}_{-7.6}$\\
$\rm r_{h}$(exponential)(arcmin) & $\rm 6.2\pm{1.0}$& $\rm 4.7\pm{1.0}$& $\rm 8.1\pm{2.7}$& $\rm 4.7\pm{0.7}$& $\rm 8.6\pm{2.5}$\\
$\rm r_{h}$(Plummer)(arcmin) & $\rm 5.4\pm{0.8}$& $\rm 4.3\pm{0.8}$& $\rm 8.3\pm{2.2}$& $\rm 4.8\pm{0.5}$& $\rm 8.8\pm{1.9}$\\
$\rm r_{h,g}$(exponential)(pc) & $\rm 24.9\pm{4.0}$& $\rm 18.9\pm{4.0}$& $\rm 196.0\pm{63}$& $\rm 32.8\pm{4.9}$& $\rm 189.8\pm{55}$\\
$\rm r_{h,g}$(Plummer)(pc) & $\rm 21.7\pm{3.2}$& $\rm 17.3\pm{3.2}$& $\rm 200.8\pm{53}$& $\rm 33.5\pm{3.5}$& $\rm 194.2\pm{42}$\\
\textbf{Background level($\rm arcmin^{-2}$)}& \textbf{0.11} &
\textbf{0.1} & \textbf{0.02} & \textbf{0.12} &
\textbf{0.12}\\
$\rm M_{V}$(exponential)(mag) & -0.77& -1.63& -4.15& -2.99& -3.91\\
$\rm M_{V}$(Plummer)(mag) & -0.81& -1.42& -4.16& -2.98& -3.92\\
 \hline
\end{tabular}\end{center}
\end{table*}



\label{lastpage}

\end{document}